\documentclass[final]{ws-ijmpa}
\def\draftnote{}
\def\catchline#1#2#3#4#5{}
\usepackage{amsmath,amssymb}
\usepackage{epsfig}
\usepackage{hyperref}

\def\as{\ensuremath{\alpha_{s}}}

\def\ab{\ensuremath{\alpha_{b}}}
\def\al{\ensuremath{\alpha_{\tau}}}
\def\at{\ensuremath{\alpha_{t}}}

\def\scale{\ensuremath{\bar\mu}}
\def\scaledec{\ensuremath{\mu_{\mathrm{dec}}}}

\def\DR{\ensuremath{\overline{\mathrm{DR}}}}
\def\MSbar{\ensuremath{\overline{\mathrm{MS}}}}
\def\MS{\ensuremath{{\mathrm{MS}}}}

\def\mtdr{\ensuremath{m_t^{\DR}}}
\def\mbdr{\ensuremath{m_b^{\DR}}}
\def\barmb{\ensuremath{\bar{m}_b}} 
\def\mbms{\ensuremath{m_b^{\MSbar}}}
\def\mlms{\ensuremath{m_\tau^{\MSbar}}}
\def\mldr{\ensuremath{m_\tau^{\DR}}}
\def\Ml{\ensuremath{M_\tau}}
\def\Mt{\ensuremath{M_t}}
\def\Mb{\ensuremath{M_b}}

\def\MeV{\mbox{MeV}}
\def\GeV{\mbox{GeV}~}
\def\TeV{\mbox{TeV}}

\def\Mew{\ensuremath{M_Z}}

\def\L{\ensuremath{ {\mathcal L} } }

\def\mmAo{\ensuremath{M^2_{A}}}
\def\mmho{\ensuremath{M^2_{h}}}
\def\mmGo{\ensuremath{M^2_{G_0}}}
\def\mmHo{\ensuremath{M^2_{H}}}
\def\mmHp{\ensuremath{M^2_{H^\pm}}}
\def\mmGp{\ensuremath{M^2_{G^\pm}}}
\def\mmQ{\ensuremath{M^2_{\tilde Q}}}
\def\mmU{\ensuremath{M^2_{\tilde t}}}
\def\mmD{\ensuremath{M^2_{\tilde b}}}

\newcommand{\mmsb}[1]{\ensuremath{m^2_{\tilde b_{#1}}}}
\newcommand{\mmst}[1]{\ensuremath{m^2_{\tilde t_{#1}}}}
\newcommand{\mmsl}[1]{\ensuremath{m^2_{\tilde \tau_{#1}}}}

\def\mgl{\ensuremath{m_{\tilde g}}}
\def\mmgl{\ensuremath{m^2_{\tilde g}}}

\def\mmsnl{\ensuremath{m^2_{\tilde \nu}}}

\def\CP{\ensuremath{\mathcal{CP}}}
\newcommand{\dzeta}[2]{ \ensuremath{ \delta \zeta_{#1}^{(#2)} } }
\newcommand{\oeps}[1]{\ensuremath{\frac{#1}{\varepsilon}}}

\begin{document}
\markboth{A.V.~Bednyakov}
{On the two-loop decoupling corrections to $\tau$-lepton and $b$-quark running masses in the MSSM}

%
\catchline{}{}{}{}{}
%

\title{ON THE TWO-LOOP DECOUPLING CORRECTIONS \\ TO $\tau$-LEPTON AND $b$-QUARK RUNNING MASSES IN THE MSSM}
\author{A.V.~BEDNYAKOV}
\address{BLTP, Joint Institute for Nuclear Research, Dubna, Russia\\
bednya@theor.jinr.ru}

\date{\today}
\maketitle

\begin{abstract}
	Masses of heavy Standard Model fermions (top-quark, bottom-quark, and tau-lepton) play an important
	role in the analysis of theories beyond the SM.
	They serve as low-energy input and reduce the parameter space of such theories.	
	In this paper Minimal supersymmetric extension of the SM is considered and 
	two-loop relations between known SM values of fermion masses 
	and running parameters of the MSSM are studied within the effective theory approach. 
	Both $b$-quark and $\tau$-lepton have the same quantum numbers with respect to $SU(2)$ group	
	and in the MSSM acquire their masses due to interactions with the same Higgs doublet.  	
	As a consequence, for large values of $\tan\beta$ parameter corresponding Yukawa couplings
	also become large and together with $\tan\beta$ can significantly enhance radiative corrections.
	In the case of $b$-quark two-loop $\mathcal{O}(\alpha_s^2)$ contribution to the relation 
	between running bottom-quark mass in QCD and MSSM is known in literature.
	This paper is devoted to calculation of the NNLO corrections proportional to Yukawa couplings. 
	For the $\tau$-lepton obtained contribution can be considered as a good approximation to the full two-loop
	result. For the $b$-quark numerical analysis given in the paper shows that 
	only the sum of strong and Yukawa corrections can play such a role.
\end{abstract}

\ccode{PACS numbers: 12.38.Bx, 12.60Jv, 14.65Fy, 14.60Fg}

\section{Introduction \label{sec:intro}}
One of the remarkable properties of the supersymmetric (SUSY) extensions 
of the Standard Model is the possibility to obtain nice unification
	of gauge couplings at the GUT scale. 
By means of one-loop renormalization group analysis of the Minimal Supersymmetric Standard Model (MSSM)
	in the beginning of 90s of the last century 
	the scale of SUSY breaking (\TeV) compatible with the 
	unification at $10^{16}$ \GeV was ``predicted'' \cite{Amaldi:1991zx}. 
At present modern computer codes 
	(SOFTSUSY \cite{Allanach:2001kg}, SuSpect \cite{Paige:2003mg}, SPheno \cite{Porod:2003um},
	ffmssmsc \cite{ffmssmsc})
	routinely use two-loop renormalization group equations (RGE) to calculate
	the spectrum of superparticles given high energy input for SUSY breaking parameters.
It is obvious that RGE at higher loops become a system of coupled differential equations so
	even to study gauge coupling unification one needs to know the value of other, e.g. Yukawa, couplings.

Since for the moment the mass of an elementary particle is the only source of information
	about its coupling to Higgs boson(s), fermion masses are  
	important low-energy input for all the models beyond the SM.

Top quark, bottom quark and tau-lepton are considered to be the heaviest fermions known in nature.
In the context of the MSSM heavy ($b$ and $t$) quark masses 
	were studied in literature and leading two-loop relations between pole 
	and running masses were found\cite{Pierce:1992hg,Bednyakov:2002sf,Bednyakov:2004gr}. 

The pole mass  does not depend on the renormalization scale $\scale$ \cite{Tarrach:1980up,Kronfeld:1998di} 
	so the running masses can 
	in principle be extracted\footnote{
One should keep in mind that we are talking about experimental constraints on running parameters of a model that
allows one to reduce the parameter space, expressing, e.g., running masses of fermions in terms of
	of other parameters (heavy particle masses)}  
 from it at any value of $\scale$. 
However, in practice the scale should be tuned in a proper way to avoid large (high-order) radiative corrections. 

At the electroweak scale ($\sim \Mew$) the top-quark pole mass $\Mt$ can be used to find the value of the running mass 
	$\mtdr$ 
	defined in \DR-renormalization scheme\cite{Siegel:1979wq,Siegel:1980qs,Stockinger:2005gx}.  
This is due to the fact that $\Mew \sim M_t$ and there
	are no large logarithms in the relation.

The story becomes more involved if one considers $b$-quark and $\tau$-lepton. 
In a theory with many different mass scales $m \ll M$ it is not so easy to avoid the appearance of  
	large contributions in the form of $\log M/\scale$ and $\log m/\scale$ 
	with $m, M$ corresponding to masses of light and heavy particles.  
	This is a ``non-decoupling feature'' of minimal (\MS-like) renormalization schemes \cite{Appelquist:1974tg}.
In our case we have $\Mb,\Ml \ll \Mew$ and there are large logarithms in the relations. 
Moreover, for $b$-quark there is a renormalon ambiguity\cite{Beneke:1994sw} 
	that limits the precision of experimental pole mass determination.

In order to solve the problem one usually employ the 
	concept of effective field theory (see, e.g., Ref.~\refcite{Georgi:1994qn} for a review)
	and perform ``manual'' decoupling of heavy particles. 
The procedure is well-known in the context of QCD~\cite{Bernreuther:1981sg} 
	(see also a nice program RunDec\cite{Chetyrkin:2000yt}). 

This approach allows one to relate running parameters in well-established effective theory 
	and corresponding 
	parameters in a more fundamental theory by means of so-called decoupling constants
	which can be calculated order by order in perturbation theory. 
In some sense decoupling constants absorb 
	leading contribution of heavy particles to various low-energy quantities. 
For the $b$-quark two-loop $\mathcal{O}(\as^2)$ contribution  
	to the relation between \mbdr~and \mbms~due to strong interactions was obtained within MSSM 
	in Refs.~\refcite{Bednyakov:2007vm,Bauer:2008bj}.
It is known from one-loop\cite{Pierce:1996zz} and two-loop\cite{Bednyakov:2004gr} calculations
	that strong corrections with virtual supersymmetric particles can be significantly 
	reduced by contributions due to other interactions. 
In this paper the results for two-loop corrections proportional to Yukawa couplings of heavy 
	SM fermions $\alpha_f = y_f^2/(4\pi)$ with $y_f = \{y_t, y_b, y_\tau\}$ will be presented. 

Contrary to the quark masses leptons do not have large uncertainties due to confinement 
	so tau-lepton pole mass can be extracted from the experiment with high precision
	$\Ml = 1776.84\pm0.17~\MeV$ (see e.g. Ref.~\refcite{Amsler:2008zzb}). 

In principle this fact allows one to determine the value of the running mass (or equivalently 
	Yukawa coupling) very precisely.  
In the MSSM only one-loop supersymmetric contribution to the relation between pole \Ml~ and running \mldr~ 
	masses is known\cite{Pierce:1996zz}.
The value of the correction depends on parameters of the MSSM and in some cases
	can be of the order of 10\%. 

Clearly, in comparison with the experimental uncertainty one-loop contribution is rather big.
So it seems to be a good idea to calculate two-loop corrections. 
Moreover, a general reasoning tells us that the inclusion of the two-loop result allows one to 
	reduce the dependence of the final result 
	on the renormalization (or decoupling) scale $\scale$. 

Why decoupling constants are important in studying a theory beyond the SM? 
Together with renormalization group equations (RGE) 
	they allow one to use a power of \MS-like schemes in studying high-energy behavior of the MSSM. 
According to formal perturbation theory in order to obtain the value of, e.g., MSSM $b$-quark running Yukawa coupling 
	$y_b$ at the GUT scale with $L$-loop precision one needs to perform a matching
	of an effective theory (e.g., SM or even Fermi theory) with more fundamental MSSM 
	at $(L-1)$-loop level somewhere at the electroweak (or SUSY) scale\footnote{
In principle, the result does not depend on the decoupling scale. 
Again due to truncation of the perturbative series one has to be careful when choosing a particular value.}.
So for one-loop RGE analysis decoupling constants are trivial and running 
	parameters are continuous when one crosses a threshold of some heavy particle.
The situation becomes more involved when two- or three-loop RGEs are employed. 
The parameters obtain a non-zero shift at the scale at which a heavy particle is decoupled.
As it was mentioned above many codes use two-loop RGEs and one-loop decoupling corrections are incorporated. 
It should be pointed out that there exist a dilemma at which scale to decouple heavy particles and how many particles
	to decouple at chosen scale. 

The problem is that when we cross the threshold and decouple a particle we sometimes break a symmetry
	that guarantees the equality of coupling constants that enter different interaction vertices
	in a Lagrangian. 
For example, decoupling of only one squark breaks the supersymmetry which relates the interaction
	of quarks, gluino and squarks to that of quarks and gluons. 
As a consequence, one needs to introduce a new coupling constant in the effective theory without the squark. 
This coupling coincide with the strong coupling constant $g_s$ above the decoupling scale $\scaledec$
	but is not equal to $g_s$ below $\scaledec$. 
Of course, the difference can be calculated. 
However, when one goes from the MSSM to the SM and decouples every heavy superparticle at its mass
	a bunch of intermediate effective theories are produced with different symmetries broken 
	(it can also be $SU(2)$ symmetry of the SM) with different RGE equations and different threshold
	corrections.
This way is certainly not the optimal one. 
In order to make use of all the symmetries presented in full theory during calculation of
	threshold corrections it seems to be a good idea to match the SM directly to the MSSM (``common scale approach'')
	\cite{Baer:2005pv}.
So we are left with the issue of choosing the decoupling scale. 
A common choice is \Mew~scale. 
	(see, e.g.~Refs.~\refcite{Allanach:2001kg,Baer:2005pv}). 
Clearly, at \Mew~high order terms can become important. 
Moreover, there exist some MSSM scenarios\cite{Giudice:2004tc} 
	when masses of scalar superparticles significantly differ
	from that of fermion superparticles. 
In this case high order corrections can also improve the 
	precision of calculation.	 

In fact, in the context of MSSM RGEs are known up to three loops\cite{Jack:2004ch},.  
The analysis presented in Ref.~\refcite{Jack:2004ch} was based on  
	one-loop threshold (decoupling) corrections and it was mentioned the necessity 
	of two-loop results for self-consistent study.
The results presented in this paper together with that obtained earlier\cite{Bednyakov:2007vm,Harlander:2007wh} 
	are aimed to partially fix this mismatch.
	
The paper organized as follows. 
First of all, the approximation to the MSSM (so-called gauge-less limit) is described in Section~\ref{sec:gaugeless}. 
A special attention is paid to the tadpole diagram treatment in Section~\ref{sec:tadpoles}.
Then a brief review (see Sec.~\ref{sec:details_calc}) of decoupling procedure is presented .
Section~\ref{sec:results} is devoted to the results 
	and numerical analysis of the calculated contribution in a wide range of parameter space of the MSSM.
In the end of the paper Conclusions and Acknowledgments can be found.

\section{Gauge-less limit of the MSSM \label{sec:gaugeless}}

In order to simplify the decoupling procedure as an effective theory I considered a theory of 
	free tau-lepton and five-flavor QCD with massive bottom-quark. 
Due to smallness of electroweak gauge coupling I neglected them. 
Moreover, the lightest Higgs boson is assumed to be much heavier then 
	the bottom quark and tau-lepton so it is ''left'' in the MSSM. 

In the effective theory we employ  $\MSbar$ minimal renormalization scheme and 
	have the following set of running parameters:
	mass of the tau-lepton\footnote{In the considered approximation ``running mass'' coincide with
	the pole mass \Ml.} (\mlms), mass of the $b$-quark (\mbms) and strong coupling constant
	($g_s^{\MSbar}$). 
By means of well-known technique \cite{Bernreuther:1981sg} they can be related to the parameters of the MSSM
	defined in so-called \DR-scheme\footnote{The issue of $\MSbar \to \DR$ transition can be
	solved be decoupling of unphysical $\varepsilon$-scalars in a way presented in Ref.~\refcite{Bednyakov:2007vm}
	or by a two-step procedure given  in Ref.~\refcite{Harlander:2007wh}}.

In such a simple setup decoupling corrections to the tau-lepton mass coincide with the corrections\footnote{Actually, with the first term of Large Mass Expansion\cite{Tkachov:1983pa,Tkachov:1984us,Smirnov:2002pj}} that
	enter the relation between the \mldr~and the pole mass \Ml. 
Due to strong interactions for the $b$-quark we have to take into account the difference between \mbms~and \Mb.

Since I neglected electroweak gauge interactions in the effective theory it is convenient to do the same thing
	in the MSSM. This approximation is called a gauge-less limit of the MSSM\cite{Haestier:2005ja}.    

In this limit there is no mixing between gauginos and higgsinos so only the latter have to be taken into account. 
Both charged and neutral higgsinos have the same mass that is equal to the absolute value of 
	the supersymmetric Higgs mixing parameter $\mu$ \footnote{
In this work I assumed that $\mu>0$ which corresponds the positive contribution to the muon anomalous magnetic 
	moment.}.
The mixing matrices for chargino ($U$ and $V$) and neutralino ($N$) are given by the following expressions
\begin{align}
	U = V = \begin{pmatrix} 1 & 0 \\ 0 & 1 \end{pmatrix},
	\qquad N = \begin{pmatrix} 1 & 0 & 0 & 0 \\
                                  0 & 1 & 0 & 0 \\
				  0 & 0 & \frac{i}{\sqrt 2} & \frac{i}{\sqrt 2} \\
				  0 & 0 & - \frac{1}{\sqrt 2} & \frac{1}{\sqrt 2}
		   \end{pmatrix}
\label{higgsino:no_mixing}
\end{align}

It should be mentioned that in Higgs sector we can have a problem since 
in the MSSM quartic interaction of Higgs bosons are proportional 
to the electroweak gauge couplings.
Nevertheless, in this paper I assume that there is a successful electroweak symmetry breaking so Higgs bosons 
	have non-trivial vacuum expectation values and treat the gauge-less limit in a formal way.

It is fair to say, that I am not going to be completely self-consistent within the limit. 
In the next section, Higgs sector of the gauge-less version of MSSM will be discussed 
together with an issue related to so-called tadpole diagrams. 
As it will be shown, in our case we have a very degenerate situation in Higgs sector with four massless bosons
	and four bosons with equal masses.
Clearly, this is not satisfactory from the phenomenological point of view. 

\section{Higgs sector and tadpoles\label{sec:tadpoles}}

In the MSSM like in the SM the electroweak symmetry is broken by a vacuum state, which can be characterized
	by the vacuum expectation values $v_1$ and $v_2$ of two Higgs doublets $H_1$ and $H_2$  
\begin{equation}
H_1= \frac{1}{\sqrt{2}}
    \begin{pmatrix} v_1 + \phi_1-i \chi_1 \\
               -\sqrt{2} \phi_1^{-} \end{pmatrix}, \quad
H_2= \frac{1}{\sqrt{2}}
    \begin{pmatrix} \sqrt{2} \phi_2^{+} \\
                    v_2+\phi_2+i\chi_2
    \end{pmatrix}.
\label{HiggsFields}
\end{equation}

In the true vacuum the first derivative of the effective potential (tadpole) should vanish.  
	However, naive calculation of one-loop tadpoles for neutral \CP-even
    	Higgs bosons $h$ and $H$  shows that they do not vanish. 
This can be interpreted in the following way. 
The vacuum expectation values for the Higgs bosons that we substituted
 	in the tree-level Higgs potential minimize the tree-level potential instead of effective one.

In order to deal with the problem I adopt the reasoning of Refs.~\refcite{Pierce:1992hg,Pierce:1996zz}. 
Tree-level tadpoles for $\phi_1$ and $\phi_2$ that appear in the MSSM Lagrangian after substitution 
of \eqref{HiggsFields} look like
\begin{eqnarray}
	T_1 & = & \frac{g^2 + g'^2}{8} (v_1^2 - v_2^2) v_1 + m_1^2 v_1 - m_3^2 v_2,  \nonumber\\
	T_2 & = & \frac{g^2 + g'^2}{8} (v_2^2 - v_1^2) v_2 + m_2^2 v_2 - m_3^2 v_1, 
\label{higgs:tadpoles}
\end{eqnarray} 
	where $g$, $g'$ are gauge $SU(2)\times U(1)$ couplings (they will be neglected in what follows) 
	and $m^2_1,m^2_2$, and $m^2_3$ are 
	soft supersymmetry breaking parameters of the Higgs potential\cite{Pierce:1992hg}.
Both $T_1$ and $T_2$ 
	are assumed to be non-zero and serve as counter-terms to cancel loop-induced tadpoles. 
As a consequence, I do not need to consider diagrams with tadpole insertions like in 
	Refs.~\refcite{Hempfling:1994ar,Jegerlehner:2003py,Faisst:2004gn} 
	since they are precisely canceled by the tree-level counter-term.  
However, they do not disappear completely 
	and manifest themselves in the mass matrices of all Higgs bosons. 
In the gauge-less limit mass matrices have the following form ($\Phi = \{\phi_i, \phi^\pm_i, \xi_i\}$, $i=1,2$)
\begin{eqnarray}
	M_\Phi  =   \begin{pmatrix} m_1^2 & - m_3^2 \\ - m_3^2 & m_2^2 \end{pmatrix}
	& = & m^2_3 \begin{pmatrix} \cot \beta & -1 \\
				-1 & \tan \beta \end{pmatrix} + \delta M_\Phi 
\label{higgs:mass_matrix_gaugeless} \\
	\delta M_\Phi & = &
		\begin{pmatrix}
				\frac{T_1}{v_1} & 0 \\
				0 & \frac{T_2}{v_2}	
		\end{pmatrix}  
\label{higgs:mass_matrix_tadpole_contrib} 
\end{eqnarray} 

Since $T_1$ and $T_2$ should cancel loop-induced tadpoles they are at least $\mathcal{O}(g^2)$ with $g$ 
	being some coupling constant of the theory. 
Consequently, in order to obtain tree-level mass matrices one should set $T_1=T_2 =0$
	and diagonalize only first term in \eqref{higgs:mass_matrix_gaugeless}. 
Clearly, this matrix has two eigenvalues $(0, m_3^2)$ and it is diagonalized by the rotation with angle $\beta$.

As it was mentioned in the end of the previous section this is not a satisfactory result. 
I decided to be slightly more close to the MSSM and introduce 
	different masses for all physical Higgs bosons together with a different mixing angle
	$\alpha$ for $\CP$-even states 
\begin{eqnarray}
	\begin{pmatrix}
		H \\ h 
	\end{pmatrix}
	  & = & R(\alpha) 
	 \begin{pmatrix}
	 	\phi_1^0 \\ \phi_2^0
	 \end{pmatrix}, \qquad 
		R(\theta) = \begin{pmatrix} \cos \theta & \sin \theta \\ -\sin\theta & \cos \theta\end{pmatrix}.
		\label{mixing:Rmatix}
\label{mixing:CP-even-higgs} \\
	\begin{pmatrix}
		G_0 \\ A 
	\end{pmatrix}
	 & = & 
	 R(\beta)
	 \begin{pmatrix}
	 	\chi_1^0 \\ \chi_2^0
	 \end{pmatrix},
\label{mixing:CP-odd-higgs} \\
	\begin{pmatrix}
		G^+ \\ H^+ 
	\end{pmatrix}
	 & = & 
	 R(\beta)
	 \begin{pmatrix}
	 	\phi_1^+ \\ \phi_2^+
	 \end{pmatrix}.
\label{mixing:charged_higgs} 
\end{eqnarray}
	After tree-level diagonalization linear and quadratic parts of the Higgs potential 
	can be rewritten in the following way:
\begin{eqnarray}
	V_2 & = & T_H \, H + T_h \, h 
	    + \frac{1}{2} \left( M_H^2 \, H^2 + M_h^2 \, h^2 + M_A^2 \, A^2 \right)
	    + M_{H^\pm}^2 H^+ H^- \nonumber \\
	 & +  & \begin{pmatrix}
	 	H & h  
		\end{pmatrix}
		\begin{pmatrix} 
			b_{HH} & b_{hH} 
			\\ b_{hH} & b_{hh} 
		\end{pmatrix}
		\begin{pmatrix}
		 H \\ h 
		\end{pmatrix} 
 	   +    \begin{pmatrix} G^0 & A \end{pmatrix} 
			\begin{pmatrix} b_{G^0G^0} & b_{G^0A} \\ b_{G^0A} & b_{AA} \end{pmatrix} 
	  		\begin{pmatrix} G^0 \\ A \end{pmatrix} \nonumber \\
	&  + &	 \begin{pmatrix} G^+ & H^+ \end{pmatrix} 
			\begin{pmatrix} b_{G^+G^-} & b_{G^+H^-} \\ b_{G^+H^-} & b_{H^+H^-} \end{pmatrix} 
		\begin{pmatrix} G^- \\ H^- \end{pmatrix}.
\end{eqnarray}
Here $H$, $h$ are neutral \CP-even Higgs bosons, $A$ and $G^0$ --- neutral \CP-odd higgs 
	and 
	Goldstone boson correspondingly, $H^\pm$ and $G^\pm$ --- charged higgs and 
	goldstone bosons.
	Additional contributions to mass matrices come from $\delta M_\Phi$ 
	(see Eq.~\eqref{higgs:mass_matrix_tadpole_contrib}) and
	can be expressed in terms
	of rotated tadpoles $T_H,~T_h$
\begin{equation}
	\begin{pmatrix} T_H \\ T_h \end{pmatrix}
= 
	R(\alpha) \begin{pmatrix} T_1 \\ T_2 \end{pmatrix},
\label{tadpoles_higgs_rotated}
\end{equation}
	a vacuum expectation value $v = \sqrt{v_1^2+v_2^2}$, and Higgs mixing angles $\beta, \alpha$ 
	\cite{Pierce:1992hg} 
\begin{subequations}
\label{TadMass}
\begin{eqnarray}
b_{HH}& = & \frac{2}{v \, s_{2\beta}}
\left(T_H(c_\alpha^3\, s_\beta +s^3_\alpha c_\beta)\right.
\left.+T_h s_\alpha c_\alpha s_{\alpha - \beta}\right), \label{HH} \\
b_{Hh} & = & \frac{s_{2\alpha}}{v \,  s_{2\beta}}
\left(T_H s_{\alpha-\beta}
+T_h c_{\alpha-\beta}\right), \label{Hh} \\
b_{hh} & = & \frac{2}{v \, s_{2\beta}}
\left(T_H c_\alpha s_\alpha c_{\alpha-\beta}\right.
\left.+T_h( c^3_\alpha c_\beta
-s^3_\alpha s_\beta)\right), \label{hh} \\
b_{G^0G^0}& = &  b_{G^+ G^-} = \frac{1}{v}\left(T_H c_{\alpha-\beta} -T_h s_{\alpha-\beta}\right), \label{GG} \\
b_{G^0A}& = &  b_{G^+ H^-} =  \frac{1}{v}\left(T_H s_{\alpha-\beta} +T_h c_{\alpha-\beta}\right), \label{GA} \\
b_{AA} & = & b_{H^+ H^-}  = \frac{2}{v \, s_{2\beta}} 
\left(T_H( s^3_\beta c_\alpha +c^3_\beta s_\alpha)+
T_h(c^3_\beta c_\alpha -s^3_\beta s_\alpha)\right), \label{AA}
\end{eqnarray}
\end{subequations}
	where $c_\theta = \cos \theta$ and $s_\theta = \sin \theta$ for some angle $\theta$.
Since the contribution due tadpoles does not depend on external momenta it can be taken into account by
	introduction of additional non-minimal counter-terms for higgs masses and mixing.  

In the end of this section I would like to mention another issue related to Goldstone bosons.
In the gauge-less limit there are no gauge bosons to ``eat'' goldstones so the latter are massless and  
	the global $SU(2)\times U(1)$ symmetry prevents them to acquire mass.
However, in perturbation theory it is not so obvious. 
For example, if one does not take into account 
	the contributions \eqref{TadMass} during calculation of goldstone boson self-energies
	$\Sigma_{GG}(p^2)$ non-zero masses for $G^0$ and $G^\pm$ are immediately generated since $\Sigma_{GG}(0) \neq 0$. 
It worth mentioning that we also run into a problem with spurious infra-red (IR) divergences when we try to calculate 
	Feynman integrals with $\Sigma_{GG}(p)$ insertions.
The role of tadpoles is crucial here since they precisely cancel this non-zero contribution keeping Goldstone
	bosons massless.

In deriving the result presented here I was using non-zero masses for Goldstone bosons denoted by $\mmGo$ and $\mmGp$ 
	which allows me not to deal with
	mentioned IR problem  explicitly. This can be justified since in the full MSSM 
	when linear $R_\xi$-gauge is employed Goldstone boson masses are proportional to that of corresponding gauge
	bosons. 
	This introduces gauge dependence in the result. 
However, numerical analysis shows that the dependence is negligible if one takes into account tadpole contribution.
	 
\section{Decoupling procedure\label{sec:details_calc}}

The process of obtaining decoupling corrections for $b$-quark mass is described in great details 
	in Ref.~\refcite{Bednyakov:2007vm}. I closely follow the same procedure. 
Here I would like to stress some of important steps of the calculation.

In order to obtain a final result for the relation between running masses in the MSSM and
	our effective theory we need to know so-called ``bare'' decoupling-constants $\zeta_{m_f,B}$ 
	up to two-loop level. 
These constants relate bare parameters in the effective five-flavor QCD (underlined parameters) to that of the MSSM
	\begin{equation}
		\underline{m_{f,B}} = \zeta_{m_f,B}(\alpha_B, M_B) \times m_{f,B}, \qquad f = \{b,\tau\}
	\label{dec_relation:bare}
	\end{equation}
	and depend on bare coupling constants $\alpha_B = \{\alpha_s,\alpha_t, \alpha_b, \alpha_\tau\}_B$ 
	and on bare masses of heavy particles denoted collectively by $M_B$.

	
The needed relations can be derived from \eqref{dec_relation:bare} by proper
	renormalization of the left- and right-hand sides. 
Since $\underline{m_{f}(\scale)} \equiv m_f^{\MSbar}(\scale)$ and $m_{f}(\scale) \equiv m_f^{\DR}(\scale)$ 
	are defined in different theories we have
\begin{eqnarray}
	\underline{m_{f,B}} & = & Z^{\MSbar}_{m_f}\left(\alpha^{\MSbar}\right) 
		           \times m^{\MSbar}_f, \qquad \alpha^{\MSbar} = \alpha_s^{\MSbar}
	\nonumber\\
	m_{f,B} & = & Z^{\DR}_{m_f}\left(\alpha^{\DR}\right) \times m^{\DR}_f
\qquad
\alpha^{\DR} = \{\alpha_s,\alpha_t, \alpha_b, \alpha_\tau\}^{\DR}.
\label{bare_masses:renormalization}
\end{eqnarray}
	As a consequence 
	an implicit equation
\begin{eqnarray}
	m_f^{\MSbar} = m_f^{\DR} \times \frac{Z^{\DR}_{m_f}(\alpha^{\DR})}{Z^{\MSbar}_{m_f}(\alpha^{\MSbar})}
	\times \zeta_{m_f,B} \left(Z^{\DR}_\alpha \alpha^{\DR}, Z^{\DR}_M M^{\DR} \right)
\label{decoupling_main_formula}
\end{eqnarray}
	can be solved in perturbation theory by expressing $\alpha^{\MSbar}$ in terms of $\alpha^{\DR}$ 
	and expanding the result in $\alpha^{\DR}$.

Simple power counting tells us that in our case we need to know $Z^{\DR}_{\alpha}$ and $Z^{\DR}_M$ up
	to one-loop level and both $Z_{m_f}$'s -- up to two loops. 
Renormalization constants for the MSSM parameters $Z^{\DR}_{\alpha}$ and $Z^{\DR}_M$ can be found in~\ref{app:ren_const}.  
The expressions for $Z^{\DR}_{m_f}$ that were used in my calculation looks like
\begin{align}
	Z^{\DR}_{m_b}  =  
		 1 & - 
			   \frac{\as}{4 \pi} \left( \frac{  2 C_F  } {\varepsilon} \right)
			 + \frac{\ab}{4 \pi} \frac{   3 } { 2 \varepsilon} 
			 + \frac{\at}{4 \pi} \frac{   1 } { 2 \varepsilon} 
			   \nonumber\\
	    & -  
	    \frac{\as 
			 	       \ab 
				       }{ (4 \pi)^2 } 
			 \left(
				 \frac{ 6 C_F  }{\varepsilon^2}	
			\right) 
		         - 
			 \frac{\as 
			 	\at
				       	}{ (4 \pi)^2 } 
			 \left(
				 \frac{ 2 C_F  }{\varepsilon^2}	
			\right) 
		         + 
			 \frac{\at \ab}{ (4 \pi)^2 } 
			 \left(
				 \frac{  7 }{4 \varepsilon^2}	
			 	-  \frac{ 1 }{2 \varepsilon } 
			\right) 
			\nonumber\\
	    & +  
	    	\frac{\at^2 }{ (4 \pi)^2 } 
			 \left(
				 \frac{  13 }{8 \varepsilon^2}	
			 	-  \frac{ 5 }{ 4\varepsilon } 
			\right) 
			+ \frac{\ab^2}{ (4 \pi)^2 } 
			 \left(
				 \frac{  45  }{ 8 \varepsilon^2}	
			 	-  \frac{ 13 }{ 4 \varepsilon } 
			\right) 
	      + \frac{\ab \al}{(4 \pi)^2 }
		\left( \frac{3}{4\varepsilon^2} - \frac{3}{4 \varepsilon}
		\right)
			\nonumber\\
	    &  +  
	    \frac{\as^2}{ (4 \pi)^2 } 
			 C_F \, \left(
				   ( 2 C_F + 3 C_A - 6) \frac{ 1 }{\varepsilon^2}	
			 	 + ( 2 C_F - 3 C_A + 6) \frac{ 1 }{\varepsilon } 
			\right), 
\label{CT:bquarkmass_yukawa} \\
	Z^{\DR}_{m_\tau}  =  
		 1 & 	 + \frac{\al}{4 \pi} \frac{   3 } { 2 \varepsilon} 
	     +  
	    \frac{\al^2}{ (4 \pi)^2 } 
			 \left( \frac{33}{8\varepsilon^2} - \frac{7}{4\varepsilon} \right)
	   + 
	    \frac{\al \ab}{ (4 \pi)^2 } 
			 \left( \frac{9}{4\varepsilon^2} - \frac{9}{4\varepsilon} \right)
\label{CT:tauleptonmass_yukawa}
\end{align}
	with $C_F = 4/3$ and $C_A = 3$ being quadratic casimirs of $SU(3)$ group. 
The renormalization constants can be easily obtained, e.g., from the formula $m_f = y_f v_1/\sqrt 2$ ($f=b, \tau$)
	and well-known RGEs for Yukawa couplings $y_b$, $y_\tau$, 
	and vacuum expectation value $v_1$ (see Refs.~\refcite{Castano:1993ri,Allanach:1999mh}).

Corresponding expressions for $Z^{\MSbar}_{m_b}$ defined in five-flavor QCD can be found in 
	Refs.~\refcite{Tarrach:1980up,Fleischer:1998dw}.  
As it was mentioned earlier $Z^{\MSbar}_{m_\tau} = 1$ in our approximation.

Since there is $\alpha^{\MSbar}_s$ in the right-hand side of \eqref{decoupling_main_formula} we also need one-loop
	relation
	between $\alpha^{\MSbar}_s$ and $\alpha^{\DR}_s$.
This relation together with the expressions for full MSSM one-loop threshold corrections can be found in 
	Appendix D of Ref.~\refcite{Pierce:1996zz}. 
In order to derive needed $\mathcal{O}(\alpha_f)$ contribution to the fermion mass decoupling constants 
	self-energies presented in this reference have to be expanded 
	in external momenta and masses of the considered fermion and the mixing \eqref{higgsino:no_mixing}
	has to be taken into account. 

We are left with the only missing piece of the formula \eqref{decoupling_main_formula}, i.e., two-loop contribution 
	$\dzeta{m_f}{2}$ to $\zeta_{m_f}$. Corrections of the order $\mathcal{O}(\alpha_s^2)$ were found in
	Refs.~\refcite{Bednyakov:2007vm,Bauer:2008bj}. 
The contribution proportional to Yukawa couplings of heavy SM fermions is obtained in this paper by means
	of a FORM\cite{Vermaseren:2000nd} program specially written for calculation of decoupling constants in the MSSM.
Corresponding diagrams were generated with the help of FeynArts\cite{FeynArts,Hahn:2001rv}.

\begin{figure}[ht]
\begin{center}
\begin{tabular}{cc}
\includegraphics{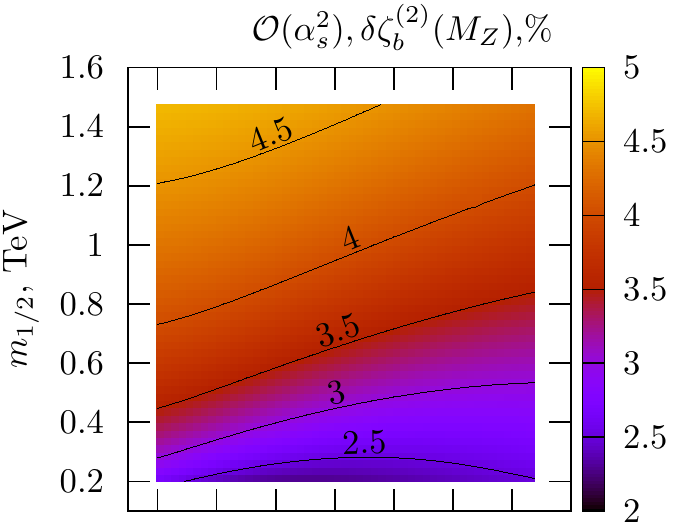} & 
\includegraphics{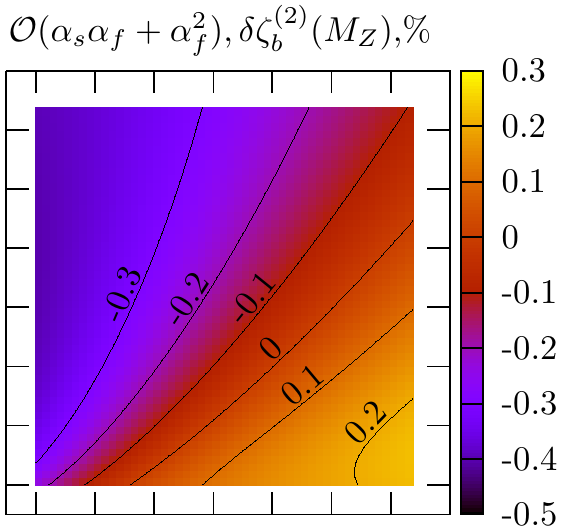} \\
\includegraphics{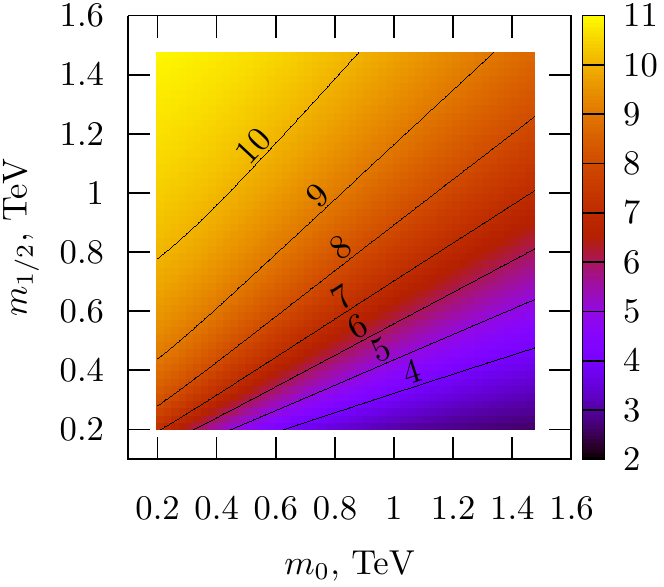} & 
\includegraphics{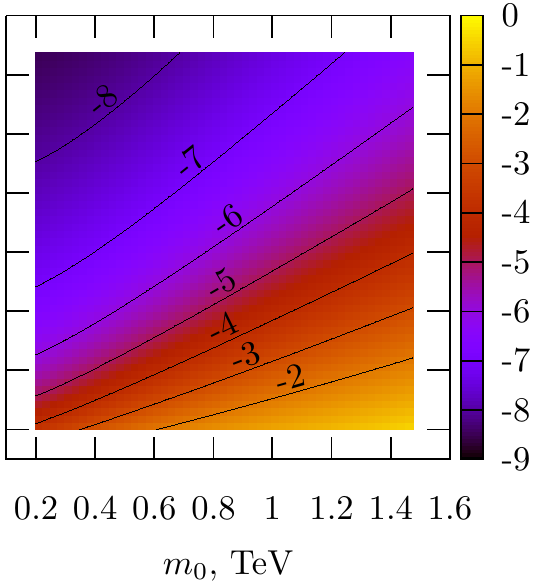} \\
\end{tabular}
\end{center}
\caption{Two-loop $\mathcal{O}(\alpha_s^2)$ and $\mathcal{O}(\alpha_s \alpha_f + \alpha_f^2)$ contributions to $\dzeta{m_b}{2}$ as functions of $m_0$ and $m_{1/2}$ for $A_0 = 0$.  
Upper row is for $\tan \beta = 10$ and the lower one is for $\tan \beta = 50$.}
\label{fig:b_strong_vs_yukawa}
\end{figure}

\section{Results\label{sec:results}} 
	In this section I present the numerical analysis of the obtained result. 
Analytical expressions are huge and have been stored in the form of Mathematica code and GiNaC 
	\cite{DBLP:journals/corr/cs-SC-0004015} archive format\footnote{Both the expressions are available
	from the author by demand.}. 
The former allows to obtain numerical value for $\dzeta{m_b}{2}$ and $\dzeta{m_\tau}{2}$ given
	the \emph{running parameters} of the MSSM. 
The latter gives us an opportunity to include the calculated corrections in the SoftSusy program \cite{Allanach:2001kg}
	to evaluate them and to see the influence of the two-loop thresholds on the 
	resulting spectrum and running parameters.
In SoftSusy matching performed at the electroweak scale so for numerical results it is assumed 
 that decoupling scale $\scaledec = \Mew$. 

\begin{figure}[t]
\begin{tabular}{cc}
\includegraphics{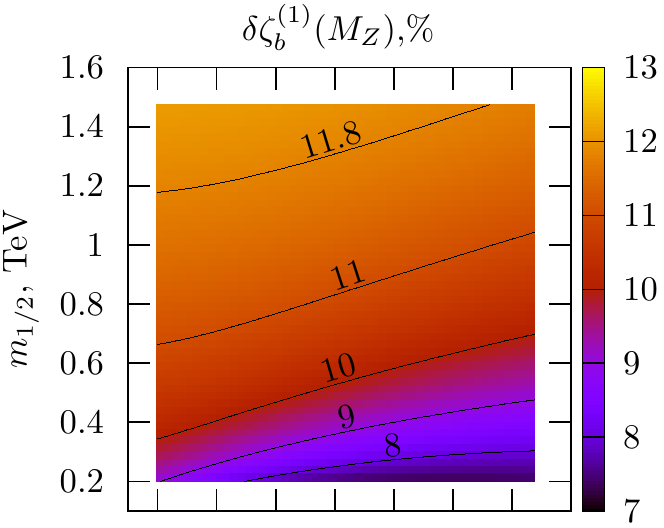} & 
\includegraphics{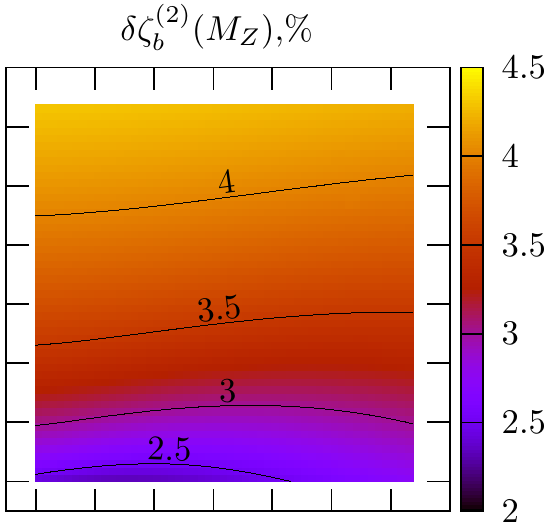} \\
\includegraphics{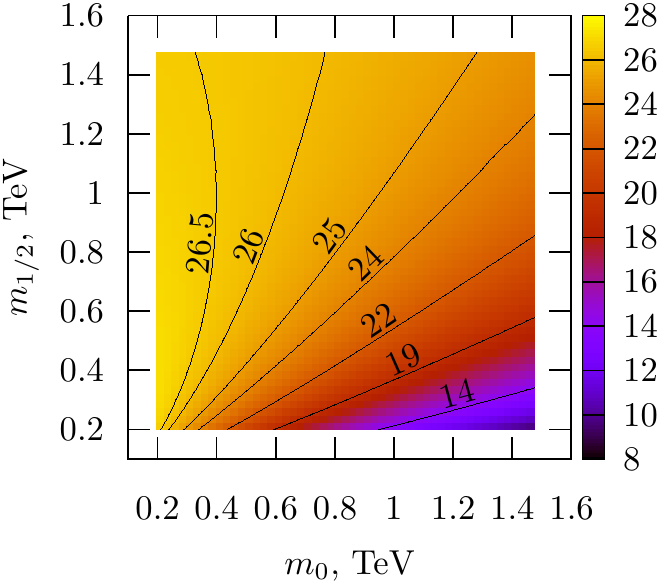} & 
\includegraphics{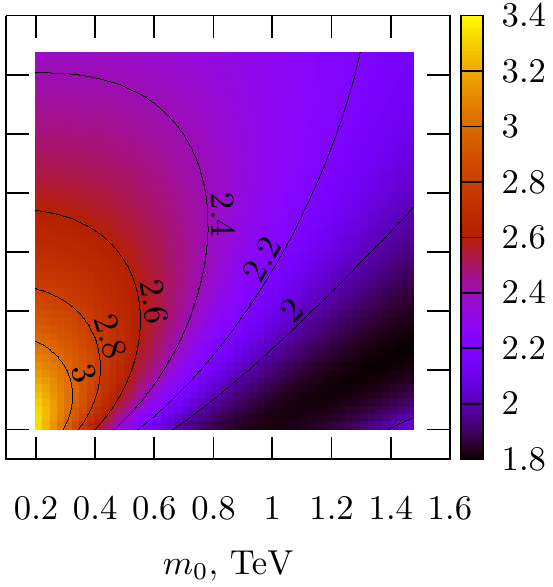} \\
\end{tabular}
\caption{Comparison of one- and two-loop contribution to the decoupling constant for the $b$-quark running mass  
$\zeta_{m_b}$ as functions of $m_0$ and $m_{1/2}$ for $A_0 = 0$.  
Upper row is for $\tan \beta = 10$ and the lower one is for $\tan \beta = 50$.}
\label{fig:b_1l_vs_2l}
\end{figure}

In Fig.~\ref{fig:b_strong_vs_yukawa} one can find typical dependence of different two-loop contributions to 
	the decoupling constant $\dzeta{m_b}{2}$ of the $b$-quark mass on Constrained MSSM  parameters $m_0$ and $m_{1/2}$ 
	for fixed value of $A_0=0$ and for $\tan\beta = 10$ (upper row) and $\tan\beta = 50$ (lower row).

As one can see corrections due to Yukawa interactions tend to compensate $\mathcal{O}(\alpha_s^2)$ contribution.
The effect of $\mathcal{O}(\alpha_s \alpha_f + \alpha_f^2)$ increases with $\tan\beta$.
The comparison of first column of Fig.~\ref{fig:b_strong_vs_yukawa} and the
	resulting two-loop corrections presented in Fig.~\ref{fig:b_1l_vs_2l}) shows us
that for large $\tan\beta = 50$ total $\dzeta{m_b}{2}$ varies in the range of 2 - 4 \% while
$\mathcal{O}(\alpha_s^2)$ contribution varies in considerable wider range  2-11 \%. 
From this fact one can immediately deduce the importance of two-loop Yukawa decoupling corrections 
for $b$-quark running mass in the region of large $\tan\beta$. 

\begin{figure}[t]
\begin{tabular}{cc}
\includegraphics{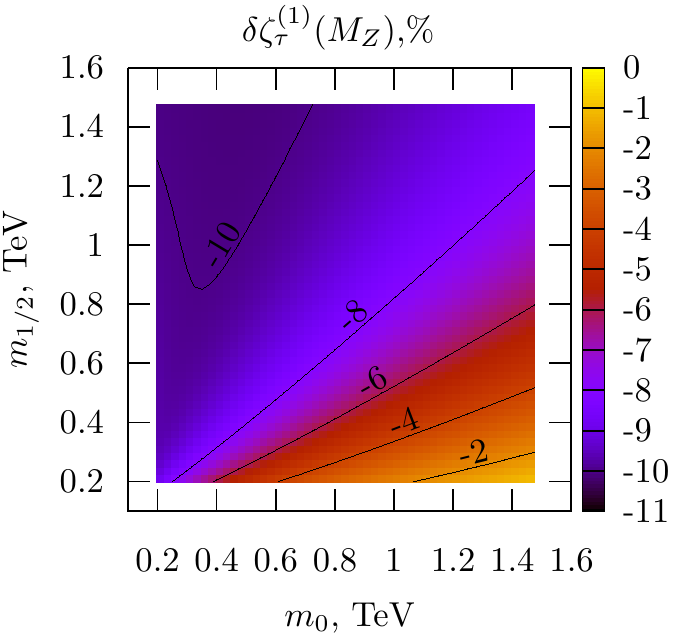} & 
\includegraphics{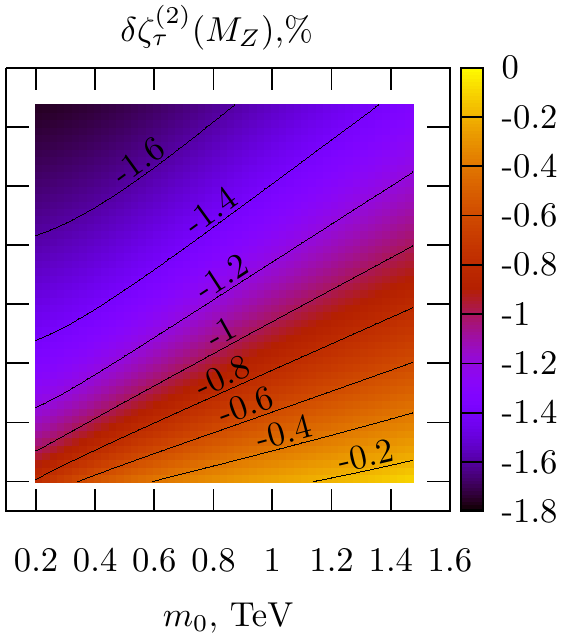} \\
\end{tabular}
\caption{Comparison of one- and two-loop contribution to $\tau$-lepton mass decoupling constant 
$\zeta_{m_\tau}$ as functions of $m_0$ and $m_{1/2}$ for $A_0 = 0$ and $\tan\beta = 50$.} 
\label{fig:tau_1l_vs_2l}
\end{figure}
 
The sum of the above contributions is compared with full one-loop MSSM decoupling constants in  Fig.~\ref{fig:b_1l_vs_2l}. 
It is easy to see that the resulting two-loop correction both for large and low values of $\tan\beta$ lies in the region
	of few percents and does not exceed the relative uncertainty of the input parameter 
	$\barmb \equiv m_b^{\MSbar}(m^{\MSbar}_b) = 4.20 \pm 0.18$ GeV \cite{Amsler:2008zzb}.

In some sense it is a bad news since we obtained the result that is negligible. 
Nevertheless, one can be sure that large one-loop threshold corrections for $b$-quark running mass
	widely discussed in literature (see, e.g., Ref.~\refcite{Carena:1999py}) are indeed reliably approximate  
	the full result.

\begin{figure}[t]
\begin{tabular}{cc}
\includegraphics[scale=0.80]{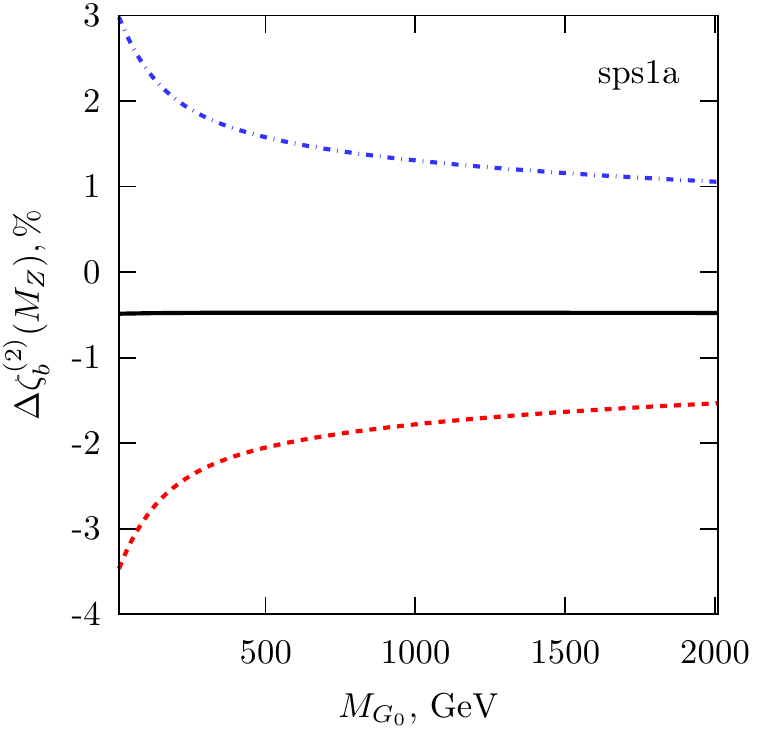} &
\includegraphics[scale=0.80]{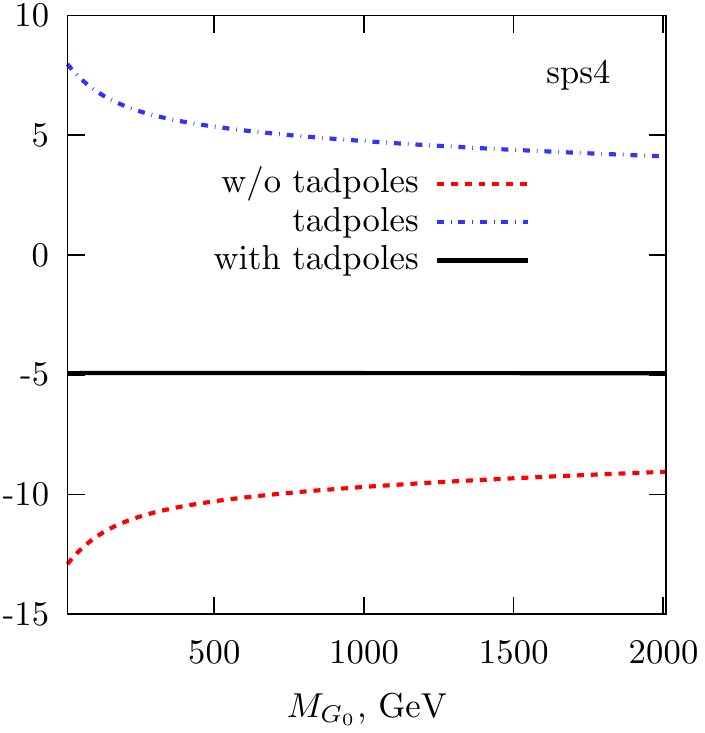}
\end{tabular}
\caption{The dependence of two-loop $\mathcal{O}(\alpha_s \alpha_f + \alpha_f^2)$ correction to $\dzeta{m_b}{2}$ on 
Goldstone boson mass $\mmGo$ for SPS1a
($m_0 = 100~\GeV$, $m_{1/2} = 250~\GeV$, $A_0 =-100~\GeV$, $\tan \beta = 10$)
and SPS4
($m_0 = 400~\GeV$, $m_{1/2} = 300~\GeV$, $A_0 = 0$, and $\tan \beta = 50$)
scenarios. Contribution due to tadpoles (dot-dashed line) almost completely cancel 
(dashed-line) naively calculated correction without tadpoles.}
\label{fig:sps1_sps4}
\end{figure}

For tau-lepton strong interactions do not contribute to the mass decoupling constant at two-loop level so
	the corrections are relatively small in this case. 
Typical value of two-loop corrections $\mathcal{O}(\al^2 + \al \alpha_q)$ with $q =t,b$ for 
	low values of $\tan\beta$ is $10^{-3}$ \% which is small even with respect to relative 
	experimental error of the pole mass $\Ml$ ($10^{-2}$ \%).
With the increase of $\tan\beta$ considered contribution is enhanced and can reach the value of few percents
	(see Fig.~\ref{fig:tau_1l_vs_2l})

In the end of this section I would like to demonstrate the role of tadpole contribution discussed 
	in Sec.~\ref{sec:tadpoles}. 
Figure~\ref{fig:sps1_sps4} shows the dependence of $\mathcal{O}(\alpha_s \alpha_f + \alpha_f^2)$
	contribution to $\dzeta{m_b}{2}$ on Goldstone boson mass $\mmGo$ for SPS1 and SPS4 scenarios \cite{Allanach:2002nj}.
One can see that if we neglect tadpole contribution we overestimate the value of the correction and introduce
	significant dependence on $\mmGo$ which can be interpreted as a gauge dependence.


\section{Conclusions\label{sec:conclusions}}

The biggest experimental facility in the world, Large Hadron Collider \cite{Evans:2008zzb}, has already been built and we are waiting	
	for first physical run of the machine.
We hope that there will be something that allow us to solve at least some of the problems of the SM and
	we believe that it will be supersymmetry.

The MSSM is a viable candidate for a theory beyond the SM. It has a lot of parameters most of which related to
	supersymmetry breaking and they have to be determined from future experiments. 
However, other parameters are already constrained from known low-energy input. 
	Both bottom-quark and tau-lepton masses are among them.

In this paper the relations between corresponding running masses $\mbdr$ and $\mldr$ defined in the MSSM 
	and known low-energy experimental input ($\barmb$ and $\Ml) $ were considered. 
Two-loop (decoupling) corrections to this relations proportional to Yukawa couplings of heavy SM fermions 
	were calculated. 

For the $b$-quark it was found that $\mathcal{O}(\alpha_s \alpha_f + \alpha_f^2)$ corrections are important
	for large $\tan\beta$ and significantly reduce two-loop strong contribution calculated earlier
	\cite{Bednyakov:2007vm,Bauer:2008bj}. 
From Figure~\ref{fig:b_strong_vs_yukawa} one can
	see that mentioned contributions usually have different signs. 
With the increase of $m_{1/2}$ ($m_0$) absolute value of the corrections increases (decreases). 
Due to this kind of behavior they tend to compensate each other in the whole region of considered $m_0 - m_{1/2}$ 
	plane lowering the sum below current uncertainty in the input parameter $\barmb$.

Two-loop threshold corrections to $\tau$-lepton mass that were obtained in this paper turns out to be negligible
	in the region of low $\tan\beta$. With the increase of $\tan\beta$ they can reach the value of 
	few percents and exceed the experimental error of $\Ml$ .

From the presented analysis it is obvious that the two-loop decoupling corrections are too small 
	and can not significantly modify MSSM spectrum produced by public computer codes. 
Variations in the spectrum due to calculated corrections are comparable with variations due to uncertainties 
	in the low-energy input parameters.

Nevertheless, obtained result allows us to be sure that, e.g. for large $\tan\beta$ 
	one-loop approximation is good enough. 
Moreover, if some new physics is  established at LHC 
	we will be ready to perform precision tests of SUSY models and their GUT extensions 
	with the help of three-loop RGEs. 

\section{Acknowledgments \label{sec:acknowledgements}}
	I would like to thank A.~Sheplyakov for fruitful discussions	
	and for his computer code \cite{bubblesII} that allows me to use GiNaC in numerical analysis.  
	Financial support from  RFBR grant No.~08-02-00856 and from a Grant for Young Scientist of JINR is kindly
	acknowledged.	
\appendix

\section{Renormalization constants \label{app:ren_const}}
To obtain finite result for threshold corrections we need to rewrite bare decoupling constants 
\eqref{decoupling_main_formula} in terms of renormalized parameters of the MSSM. 
I collected all the needed counter-terms in this Appendix. In what follows the following notations are 
used: $\mgl$ corresponds to gluino mass, $A_f$ with $f = \{t, b, \tau\}$ --- soft trilinear couplings, 
	$X_t=A_t - \mu \cot \beta$, and $X_f = A_f - \mu \tan \beta$ with $f=\{b, \tau\}$ are off-diagonal
	elements of sfermion mixing matrices. Sfermion masses are denoted by $m^2_{\tilde f_{1,2}}$ and $m^2_{\tilde \nu}$ 
	(sneutrino).
For sines and cosines of sfermion $\theta_{f}$ mixing angles  
	abbreviations $s_f \equiv \sin \theta_f$ and 
	$s_{nf} \equiv \sin (n \theta_f)$ ($n=2,4$), etc are used. 

For bare scalar mass $m_B^2$, bare fermion mass $M_B$, bare coupling constant $\alpha_B$, 
	and some bare mixing angle $\theta_B$ 
	corresponding counter-terms $\delta m^2$, $\delta Z_M$, $\delta Z_\alpha$, and $\delta \theta$ are defined by 
	$m_B^2 = m^2 + \delta m^2$, $M_B = (1 + \delta Z_M) M$, $\alpha_B = \alpha (1 + \delta Z_\alpha )$, 
	and $\theta_B = \theta + \delta \theta$ 
	with $m^2$, $M$, $\alpha$, and $\theta$  being running parameters in \DR~scheme.

\subsection{Scalar quarks}
\begin{align}
 (4 \pi) \, 	\delta \mmsb{1}  & =  
	C_F \oeps{\as}	
	\left(
		(\mmsb{2} - \mmsb{1}) s^2_{2b}
		- 4 \left(m_b^2 + \mmgl	
		- s_{2b}\, m_b \mgl 
		\right)
	\right) 
+ \oeps{\al} m_b X_\tau s_{2b}
\nonumber \\
	& +  
	\oeps{\at}
	c_b^2
		\left(
		A_t^2 
		+ \mmsb{1}
		+ \mmst{1} s_t^2
		+ \mmst{2} c_t^2
		+ \mmHp c^2_\beta 
		+ \mmGp s^2_\beta 
		\right) \nonumber \\
	& +  
	\oeps{\ab}
	s_b^2
		\left(
		A_b^2 
		+ \mmsb{1}
		+ \mmst{1} c_t^2
		+ \mmst{2} s_t^2
		+ \mmHp s^2_\beta 
		+ \mmGp c^2_\beta 
		\right) \nonumber \\
	& +  
	\oeps{\ab}
		\left(
		A_b^2  
		-  \mu^2
		+ \frac{1}{2} \left( \mmho s_\alpha^2 + \mmHo c_\alpha^2  
		+	\mmAo s^2_\beta
		+	\mmGo c^2_\beta \right)
		\right) \nonumber \\
	& +  
	\oeps{\ab}
		\left(
		m_b^2 c^2_b - m_t^2 s^2_b 
		+ 3 s_{2 b} m_b A_b 
		+ \mmsb{1} ( 1 + 2 s^2_{2 b})
		+ \mmsb{2} ( 1 - 2 s^2_{2 b})
		\right) \nonumber\\
	& +  
		\oeps{\alpha_t}
		\left(
	 m_b^2 s_b^2 - m_t^2 c^2_b +  m_b s_{2b} A_t 	
	\right)
	- \mu^2 \left(\oeps{\at} c_b^2 + \oeps{\ab} s_b^2 \right), 
\label{CT:mmsb1} 
\end{align}
\begin{align}
 (4 \pi) \, 	\delta \mmsb{2}  & =  
	C_F \oeps{\as}	
	\left(
		(\mmsb{2} - \mmsb{1}) s^2_{2b}
		- 4 \left(m_b^2 + \mmgl	
		+ s_{2b}\, m_b \mgl 
		\right)
	\right) 
- \oeps{\al} m_b X_\tau s_{2b}
	\nonumber \\
	& +  
	\oeps{\at}
	s_b^2
		\left(
		A_t^2 
		+ \mmsb{2}
		+ \mmst{1} s_t^2
		+ \mmst{2} c_t^2
		+ \mmHp c^2_\beta 
		+ \mmGp s^2_\beta 
		\right) \nonumber \\
	& +  
	\oeps{\ab}
	c_b^2
		\left(
		A_b^2 
		+ \mmsb{2}
		+ \mmst{1} c_t^2
		+ \mmst{2} s_t^2
		+ \mmHp s^2_\beta 
		+ \mmGp c^2_\beta 
		\right) \nonumber \\
	& +  
	\oeps{\ab}
		\left(
		A_b^2  
		-  \mu^2
		+ \frac{1}{2} \left( \mmho s_\alpha^2 + \mmHo c_\alpha^2  
		+	\mmAo s^2_\beta
		+	\mmGo c^2_\beta \right)
		\right) \nonumber \\
	& +  
	\oeps{\ab}
		\left(
		m_b^2 s^2_b - m_t^2 c^2_b 
		- 3 s_{2 b} m_b A_b 
		+ \mmsb{1} ( 1 - 2 s^2_{2 b})
		+ \mmsb{2} ( 1 + 2 s^2_{2 b})
		\right) \nonumber\\
	& +  
		\oeps{\alpha_t}
		\left(
	 m_b^2 c_b^2 - m_t^2 s^2_b -  m_b s_{2b} A_t 	
	\right)
	- \mu^2 \left(\oeps{\at} s_b^2 + \oeps{\ab} c_b^2 \right), 
\label{CT:mmsb2} 
\end{align}
\begin{align}
 (4 \pi)   \, 	\delta \theta_b  & =  \bigg[ 
	C_F \oeps{\as}
	\left(
			(\mmsb{2} - \mmsb{1})
		\frac{s_{4b}}{2}
			+ 4 m_b m_g c_{2b}		
	\right) 
	+ \oeps{\al} m_b X_\tau c_{2b}
	\nonumber \\
	& -  
	\oeps{\at} \frac{s_{2b}}{2} 
		\left(
	 	A^2_t
		- \mu^2 
		 + \mmst{1} s^2_t + \mmst{2} c^2_t 
		+ \mmHp c^2_\beta
		+ \mmGp s^2_\beta 
		- m_t^2 
		\right) \nonumber \\
	& +  
	\oeps{\ab} \frac{s_{2b}}{2} 
		 \left(
		 A_b^2 
		- \mu^2
		 + \mmst{1} c^2_t + \mmst{2} s^2_t 
		+ \mmHp s^2_\beta
		+ \mmGp c^2_\beta 
		- m_t^2 
		\right) \nonumber\\
& + 		\oeps{\alpha_b} \left(
		 (\mmsb{1} - \mmsb{2})
		s_{4b} 
		+ 3 m_b A_b c_{2b}
		\right) 
	+ \oeps{\at} m_b c_{2b} A_t 	
	\nonumber\\
 & +  
	 \frac{s_{2b}}{4} \left( \oeps{\ab} - \oeps{\at} \right) 
	\left(\mmsb{1} + \mmsb{2}  - 2 m_b^2\right)
	\bigg]
	\frac{1}{\mmsb{1}-\mmsb{2}}. 
\label{CT:thetab} 
\end{align}
	The counter-terms satisfy the following relation 
\begin{eqnarray}
	\delta \mmQ & = & \delta \mmsb{1} c_b^2 + \delta \mmsb{2} s_b^2 + (\mmsb{2} - \mmsb{1}) s_{2b} \delta \theta_b - 2 m_b \delta m_b 
	\nonumber \\
	& =  & 
	\delta \mmst{1} c_t^2 + \delta \mmst{2} s_t^2 + (\mmst{2} - \mmst{1}) s_{2t} \delta \theta_t - 2 m_t \delta m_t,
	\label{SU2relation:ct}
\end{eqnarray}
	which is a consequence of $SU(2)$-invariance of the MSSM Lagrangian.
By means of 
\begin{eqnarray}
	s_{2b} & = & \frac{2 m_b X_b}{\mmsb{1} - \mmsb{2}} \nonumber\\
	c_{2b} & = & \frac{\mmQ - \mmD}{\mmsb{1} - \mmsb{2}} \nonumber\\
\label{s2b_c2b}
\end{eqnarray}
	it is easy to convince oneself that  \eqref{CT:mmsb1} \eqref{CT:mmsb2} and \eqref{CT:thetab} can 
	be rewritten in the following way 
\begin{align}
(4\pi) \delta \mmsb{1,2} & =  
	\frac{1}{2} \bigg[ \delta \mmQ + \delta \mmD 
		\pm \frac{\mmQ - \mmD}{\mmsb{1} - \mmsb{2}} \left( \delta \mmQ - \delta \mmD \right) \nonumber\\
	&  \phantom{\frac{1}{2} \bigg[ \delta \mmQ } 
		+ 4 m_b^2 \frac{\delta m_b}{m_b}   
		\pm \frac{2 m_b X_b}{\mmsb{1} - \mmsb{2}} \left( \delta m_b X_b + m_b \delta X_b \right)
	\bigg]
\label{CT:mmsb_check} \\
(4\pi) \delta \theta_b & = 
	\frac{m_b X_b}{\left(\mmsb{1} - \mmsb{2}\right)^2} 
	\left[\left(\frac{\delta m_b}{m_b} + \frac{\delta X_b}{X_b} \right)
	      \left(\mmQ - \mmD \right) - \left( \delta \mmQ - \delta \mmD \right) \right],
\label{CT:thetab_check}
\end{align}
	where\footnote{needed counter-terms can be extracted from 
	corresponding beta-functions given, e.g., in Ref.~\refcite{Kazakov:2004mr}} 
\begin{align}
\frac{\delta m_b}{m_b} & = 
	\frac{3}{2} \oeps{\ab} + \frac{1}{2} \oeps{\at} - 2 C_F \oeps{\as}, 
\label{CT:mb_1loop} \\
\frac{\delta X_b}{X_b} & = \frac{1}{X_b} \left( \delta A_b - \delta \mu \tan \beta - \mu \delta \tan \beta \right) \\
	& = 3 \oeps{\ab} + \frac{1}{X_b} \left( 3 A_b \oeps{\ab} + A_t \oeps{\at} + X_\tau \oeps{\al} 
	+ 4 C_F \oeps{\as}\mgl  \right), 
\label{CT:ab_1loop} \\
\frac{\delta \mu}{\mu} & = 
	\frac{3}{2} \oeps{\ab+\at} + \frac{1}{2} \oeps{\al}, 
\label{CT:mu_1loop} \\
\frac{\delta \tan \beta}{\tan \beta} & = 
	- \frac{3}{2} \oeps{\at} + \frac{3}{2} \oeps{\ab} + \frac{1}{2} \oeps{\al}, 
\label{CT:tanb_1loop} \\
\delta A_b & = 
	4 C_F  \oeps{\as} \mgl + 6 A_b \oeps{\ab} + A_t \oeps{\at} + A_\tau \oeps{\al},
\label{CT:Ab_1loop} \\
\delta \mmQ & = 
	- 4 C_F  \oeps{\as} \mmgl 
	+ \oeps{\ab} \left( \mmQ + \mmU + m^2_{H_1} + A_t^2 \right) \nonumber\\
& \phantom{= - 4 C_F \mmgl \oeps{\as}  }
	+ \oeps{\at} \left( \mmQ + \mmD + m^2_{H_2} + A_b^2 \right),
\label{CT:mmQ_1loop} \\
\delta \mmD & = 
	- 4 C_F  \oeps{\as} \mmgl 
	+ 2 \oeps{\ab} \left( \mmQ + \mmU + m^2_{H_1} + A_t^2 \right),
\label{CT:mmD_1loop}
\end{align}
Renormalization constants for top-squark masses and mixing can be obtained from the expressions above
	by substitution $b \leftrightarrow t$, $c_\beta \leftrightarrow s_\beta$, 
	$c_\alpha \leftrightarrow s_\alpha$ and $\al\to0$ 
\subsection{Scalar leptons}
\begin{align}
	(4 \pi) \delta \mmsnl  & =  
		\oeps{\al}
\left(
	A_\tau^2 - \mu^2 + \mmsl{1} s^2_\tau  
	+ \mmsl{2} c^2_\tau+ \mmsnl - m_{\tau}^2 + c_\beta^2 \mmGp + s_\beta^2 \mmHp  
\right) 
\label{CT:mmnsl1}\\
 (4 \pi) \, 	\delta \mmsl{1}  & =  
	\oeps{\al}
	s_\tau^2
		\left(
		A_\tau^2 
		-  \mu^2
		+ \mmsl{1}
		+ \mmsnl
		+ \mmGp c^2_\beta 
		+ \mmHp s^2_\beta 
		\right) 
	+ \oeps{\ab}
		3 X_b m_\tau s_{2\tau}
\nonumber \\
	& +  
	\oeps{\al}
		\left(
		A_\tau^2 - \mu^2 
		+ \frac{1}{2} 
		\left(
		  \mmAo s_\beta^2
		+ \mmGo c_\beta^2 
		+ \mmho s_\alpha^2
		+ \mmHo c_\alpha^2
		\right)
		\right) \nonumber \\
	& +  
	\oeps{\al}
		\left(
		 \mmsl{1} (1 + s^2_{2\tau})
		+ \mmsl{2} (1 - s^2_{2\tau})
		+ 3 s_{2\tau} A_\tau m_\tau
		+ m_\tau^2 c^2_\tau
		\right) 
\label{CT:mmsl1}\\
 (4 \pi) \, 	\delta \mmsl{2}  & =  
	\oeps{\al}
	c_\tau^2
		\left(
		A_\tau^2 
		-  \mu^2
		+ \mmsl{2}
		+ \mmsnl
		+ \mmGp s^2_\beta 
		+ \mmHp c^2_\beta 
		\right) 
	- \oeps{\al}
		3 X_b m_\tau s_{2\tau}
\nonumber \\
	& +  
	\oeps{\al}
		\left(
		A_\tau^2 - \mu^2 
		+ \frac{1}{2} 
		\left(
		  \mmAo s_\beta^2
		+ \mmGo c_\beta^2 
		+ \mmho s_\alpha^2
		+ \mmHo c_\alpha^2
		\right)
		\right) \nonumber \\
	& +  
	\oeps{\al}
		\left(
		 \mmsl{1} (1 - s^2_{2\tau})
		+ \mmsl{2} (1 + s^2_{2\tau})
		- 3 s_{2\tau} A_\tau m_\tau
		+ m_\tau^2 s^2_\tau
		\right) 
\label{CT:mmsl2}
\end{align}
\begin{align}
 (4 \pi)   \, 	\delta \theta_\tau  & =  \bigg[ 
	 3 \oeps{\ab} m_\tau X_b c_{2l}
+ \frac{s_{2\tau}}{4} \oeps{\al}
	\left(\mmsl{1} + \mmsl{2}  - 2 m^2_\tau\right)
	\nonumber \\
	& +  
	\oeps{\al} \frac{s_{2\tau}}{2} 
		 \left(
		 A_\tau^2 
		- \mu^2
		 + \mmsnl 
		+ \mmHp s^2_\beta
		+ \mmGp c^2_\beta 
		\right) \nonumber\\
& + 		\oeps{\al} \left(
		 (\mmsl{1} - \mmsl{2})
		s_{4\tau} 
		+ 3 m_\tau A_\tau c_{2\tau}
		\right) 
	 	\bigg]
	\frac{1}{\mmsl{1}-\mmsl{2}}.
\label{CT:thetal} 
\end{align}
\subsection{Higgs sector}

In order to obtain renormalization constants for Higgs masses and mixing angles
	we need to consider divergent parts of self-energy diagrams. 
The straightforward calculation 
	leads to the following results 
\begin{align}
(4 \pi) \, \delta m^2_{\Phi_1} & =  3 \oeps{\at} 
	 s^2_{\theta} \left( \mmst{1} + \mmst{2} 
		      - 2 m_t^2 
	 + m^2_{\Phi_2} 
	+ (A_t - \mu \frac{c_\theta}{s_\theta}\,)^2 
		      \right)
	\nonumber \\
		& +  3 \oeps{\ab} 
	 c^2_\theta \left(
		\mmsb{1} + \mmsb{2} 
			- 2 m_b^2 
		+ m^2_{\Phi_2} 
	  + (A_b - \mu \frac{s_\theta}{c_\theta}\,)^2
			\right)
\nonumber\\
		& +  \oeps{\al} 
	 c^2_\theta \left(
		\mmsl{1} + \mmsl{2} 
			- 2 m_\tau^2 
		+ m^2_{\Phi_2} 
	  + (A_b - \mu \frac{s_\theta}{c_\theta}\,)^2
	\right)
\label{CT:mmPhi1}
\end{align}
\begin{align}
(4 \pi) \, \delta m^2_{\Phi_2} & =  3 \oeps{\at} 
	 c^2_{\theta} \left( \mmst{1} + \mmst{2} 
		      - 2 m_t^2 
	 + m^2_{\Phi_1}
	  + (A_t  + \mu \frac{s_\theta}{c_\theta}\, )^2
		      \right)
	\nonumber \\
		& +  3 \oeps{\ab} 
	 s^2_\theta \left(
		 \mmsb{1} + \mmsb{2}  
			- 2 m_b^2 
		 + m^2_{\Phi_1}
	+ (A_b +  \mu \frac{c_\theta}{s_\theta} \,)^2
	\right)
	\nonumber\\
		& +   \oeps{\al} 
	 s^2_\theta \left(
		 \mmsl{1} + \mmsl{2}  
			- 2 m_\tau^2 
		 + m^2_{\Phi_1}
	+ (A_b +  \mu \frac{c_\theta}{s_\theta} \,)^2
	\right)
\label{CT:mmPhi2} 
\end{align}
\begin{align}
(4 \pi) \, \delta \theta & =  \bigg[ 
		3 \oeps{\at} \left(
	  \mmst{1} + \mmst{2}  - 2 m_t^2 
	+ (A_t - \mu \frac{c_\theta}{s_\theta}\,)
	  (A_t  + \mu \frac{s_\theta}{c_\theta}\,)
		\right)
	\nonumber \\
		& -  
		3 \oeps{\ab} \left(
		\mmsb{1} + \mmsb{2} - 2 m_b^2 
	  + (A_b - \mu \frac{s_\theta}{c_\theta}\,)
	    (A_b +  \mu \frac{c_\theta}{s_\theta}\,)
		\right) \nonumber\\
		& -  
		\oeps{\al} \left(
		\mmsl{1} + \mmsl{2} - 2 m_\tau^2 
	  + (A_\tau - \mu \frac{s_\theta}{c_\theta}\,)
	    (A_\tau +  \mu \frac{c_\theta}{s_\theta}\,)
		\right) \nonumber\\
	 & +  
	\left(
	 3 \oeps{\at} -  3 \oeps{\ab}  - \oeps{\al}
	\right) 
		\left( \frac{m^2_{\Phi_1} + m^2_{\Phi_2}}{2}\right)
		\bigg]\frac{c_\theta s_\theta}{m^2_{\Phi_1} - m^2_{\Phi_2}}.
\label{CT:phi_higgs} 
\end{align}
Here $\Phi_1 = \{G,G^\pm,H \}$, $\Phi_2 = \{A,H^\pm,h\}$, $\theta = \{\beta,\beta,\alpha\}$ correspondingly.
It is easy to see that for $\delta \beta$ one obtains different expressions 
	when considers \CP-odd ($\delta\beta_0$) and charged higgses ($\delta\beta_\pm$).
This apparent problem ``is solved'' if we remember that in the true gauge-less limit 
	$\mmGo=\mmGp = 0$ and $\mmAo=\mmHp$ so $\delta\beta_0 = \delta\beta_\pm$. 
However, it is convenient to keep the masses different and instead of renormalization 
	of bare angles $\beta_0$ and $\beta_\pm$ use the following explicit counter-terms
\begin{align}
	\delta \L^{\beta}_{ct} =&  - \left(\mmGo - \mmAo\right) \delta \beta_0 \times G_0 A_0  \nonumber\\
				& - \left(\mmGp - \mmHp\right) \delta \beta_\pm \times \left( H^+ G^- + G^+ H^-\right),
	\label{CT:explict_higgs}
\end{align}
	where $\delta \beta_0$ and $\delta\beta_\pm$ are given by \eqref{CT:phi_higgs} for $(\Phi_1,\Phi_2) = (G, A_0)$
	and $(\Phi_1,\Phi_2) = (G^\pm,H^\pm)$ correspondingly.
I would like to mention that it is not the end of the story.
For the moment we neglected the contributions due to tadpoles described in Sec.~\ref{sec:tadpoles}. 
\begin{align}
(4 \pi) \, \delta_t m^2_{\Phi_1} & =  -3 \oeps{\at} 
	 s^2_{\theta} \left[ \mmst{1} + \mmst{2} 
		      - 2 m_t^2 
	+ (A_t - \mu \frac{c_\beta}{s_\beta}) 
	  (A_t - \mu \frac{c^2_\theta}{s^2_\theta} \frac{s_\beta}{c_\beta}) 
		      \right]
	\nonumber \\
		& -  3 \oeps{\ab} 
	 c^2_\theta \left[
		\mmsb{1} + \mmsb{2} 
			- 2 m_b^2 
	  + (A_b - \mu \frac{s_\beta}{c_\beta})
	    (A_b - \mu \frac{s^2_\theta}{c^2_\theta} \frac{c_\beta}{s_\beta})	
			\right]
\nonumber\\
		& -   \oeps{\al} 
	 c^2_\theta \left[
		\mmsl{1} + \mmsl{2} 
			- 2 m_\tau^2 
	  + (A_\tau - \mu \frac{s_\beta}{c_\beta})
	    (A_\tau - \mu \frac{s^2_\theta}{c^2_\theta} \frac{c_\beta}{s_\beta})	
			\right],
\label{CT:mmPhi1_tad}\\
(4 \pi) \, \delta_t m^2_{\Phi_2} & =  -3 \oeps{\at} 
	 c^2_{\theta} \left[ \mmst{1} + \mmst{2} 
		      - 2 m_t^2 
	+ (A_t - \mu \frac{c_\beta}{s_\beta}) 
	  (A_t - \mu \frac{s^2_\theta}{c^2_\theta} \frac{s_\beta}{c_\beta}) 
		      \right]
	\nonumber \\
		& -  3 \oeps{\ab} 
	 s^2_\theta \left[
		 \mmsb{1} + \mmsb{2}  
			- 2 m_b^2 
	  + (A_b - \mu \frac{s_\beta}{c_\beta})
	    (A_b - \mu \frac{c^2_\theta}{s^2_\theta} \frac{c_\beta}{s_\beta})	
	\right] 
	\nonumber\\
		& -   \oeps{\al} 
	 s^2_\theta \left[
		 \mmsl{1} + \mmsl{2}  
			- 2 m_\tau^2 
	  + (A_\tau - \mu \frac{s_\beta}{c_\beta})
	    (A_\tau - \mu \frac{c^2_\theta}{s^2_\theta} \frac{c_\beta}{s_\beta})	
	\right]
\label{CT:mmPhi2_tad} 
\end{align}
\begin{align}
(4 \pi) \, \delta_t \theta & =  -\bigg[ 
		3 \oeps{\at} \left(
	  \mmst{1} + \mmst{2}  - 2 m_t^2 
	+ (A_t - \mu \frac{c_\beta}{s_\beta}\,)
	  (A_t  + \mu \frac{s_\beta}{c_\beta}\,)
		\right)
	\nonumber \\
		& -  
		3 \oeps{\ab} \left(
		\mmsb{1} + \mmsb{2} - 2 m_b^2 
	  + (A_b - \mu \frac{s_\beta}{c_\beta}\,)
	    (A_b +  \mu \frac{c_\beta}{s_\beta}\,)
		\right) \nonumber\\
		& -  
		\oeps{\al} \left(
		\mmsl{1} + \mmsl{2} - 2 m_\tau^2 
	  + (A_\tau - \mu \frac{s_\beta}{c_\beta}\,)
	    (A_\tau +  \mu \frac{c_\beta}{s_\beta}\,)
	\right)
		\bigg]\frac{c_\theta s_\theta}{m^2_{\Phi_1} - m^2_{\Phi_2}}.
\label{CT:phi_higgs_tad} 
\end{align}
	It is interesting to note that for $\Phi_1 = \{G,G^\pm\}$ and $\Phi_2 = \{A,H^\pm\}$
\begin{align}
(4 \pi) \, \left( \delta m^2_{\Phi_1}  + \delta_t m^2_{\Phi_1} \right) & =   
	\left( 3 \oeps{\at} s^2_\beta + 3 \oeps{\ab} c^2_\beta + \oeps{\al} c^2_\beta \right) m^2_{\Phi_1},
\label{ct:mmPhi1_with_tad} \\
(4 \pi) \, \left( \delta m^2_{\Phi_2}  + \delta_t m^2_{\Phi_2} \right) & =   
	\left( 3 \oeps{\at} c^2_\beta + 3 \oeps{\ab} s^2_\beta + \oeps{\al} s^2_\beta \right) m^2_{\Phi_2} \nonumber\\
	& +  \frac{\mu}{c_\beta s_\beta} \left( A_\tau \oeps{\al} + 3 A_b \oeps{\ab} + 3 A_t \oeps{\at} \right) \nonumber\\
	& =  
	m^2_{\Phi_2} \left[ \frac{\delta \mu}{\mu} + \left( \frac{\delta v_2}{v_2} - \frac{\delta v_1}{v_1} 
	\right) \left(s_\beta^2 - c_\beta^2 \right) \right]  +   \frac{\mu}{c_\beta s_\beta} \delta B,
\label{ct:mmPhi2_with_tad} \\
(4 \pi) \, \left(\delta \beta + \delta_t \beta \right) & =   
		\left( \frac{3}{2} \oeps{\at} - \frac{3}{2} \oeps{\ab} - \frac{1}{2} \oeps{\al} \right)
	\frac{m^2_{\Phi_1} + m^2_{\Phi_2}} {m^2_{\Phi_1} - m^2_{\Phi_2}} c_\beta s_\beta \nonumber\\
	& =  - c_\beta s_\beta \left( \delta v_2 - \delta v_1 \right)
	\frac{m^2_{\Phi_1} + m^2_{\Phi_2}} {m^2_{\Phi_1} - m^2_{\Phi_2}},
\label{ct:beta_angle_vevs}
\end{align}
	where for $m_{\Phi_1}^2 = m^2_{G,G^\pm} = 0$ (true goldstones) 
	and $m^2_{\Phi_2} = m^2_A = m^2_{H^\pm} = m_3^2/(c_\beta s_\beta) \equiv B \mu/(c_\beta s_\beta)$
	the result can be cross-checked with the help of one-loop renormalization constants   	
\begin{equation}
	(4 \pi) \delta v_1  =  -\frac{3}{2} \oeps{\ab} - \frac{1}{2} \oeps{\al}, \qquad
	(4 \pi) \delta v_2  =  -\frac{3}{2} \oeps{\at},
\label{ct:v1_v2_1loop} 
\end{equation}
\begin{equation}
	(4 \pi) \delta B  =  3 A_t \oeps{\at} +  3 A_b \oeps{\ab} + A_\tau \oeps{\al}
\label{ct:B_1loop} 
\end{equation}
	and relations
\begin{equation}
	\delta \beta = \cos^2 \beta \, \delta \tan \beta \qquad  
	\delta m^2_A  
	= \delta \left( \frac{B \mu}{c_\beta s_\beta} \right)
\end{equation}
\subsection{Other renormalization constants}

For completeness I present one-loop counter-terms for couplings 
\begin{eqnarray}
	(4 \pi) \delta Z_{\as} & = & (6 - 3 C_A)  \oeps{\as} 
\label{CT:as} \\
	(4 \pi) \delta Z_{\ab} & = & - 4 C_F \oeps{\as} + 6 \oeps{\ab} + \oeps{\at} + \oeps{\al}
\label{CT:ab} \\
	(4 \pi) \delta Z_{\ab} & = & - 4 C_F \oeps{\as} + 6 \oeps{\at} + \oeps{\ab}
\label{CT:at} \\
	(4 \pi) \delta Z_{\al} & = &  3 \oeps{\at} + 4 \oeps{\al}
\label{CT:al} 
\end{eqnarray}
	and for the masses of $t$-quark $m_t$ and higgsinos $m_{\chi}$ 
	($b$-quark and $\tau$-lepton are considered in Sec.~\ref{sec:details_calc}) 	
\begin{eqnarray}
	(4 \pi) \delta Z_{m_t} & = & - 2 C_F \oeps{\as} + \frac{3}{2} \oeps{\at} + \frac{1}{2} \oeps{\ab},
\label{CT:mt} \\
	(4 \pi) \delta Z_{m_\chi} & = &  \frac{3}{2} \left(\oeps{\ab} + \oeps{\at} \right) + \frac{1}{2} \oeps{\al}
\label{CT:mchi}
\end{eqnarray}
	Clearly, $\delta Z_{m_\chi}$ coincide with the renormalization constant for $\mu$ \eqref{CT:mu_1loop}.
\providecommand{\href}[2]{#2}\begingroup\raggedright\endgroup


\begin{thebibliography}{10}

\bibitem{Amaldi:1991zx}
U.~Amaldi, W.~de~Boer, P.~H. Frampton, H.~Furstenau, and J.~T. Liu,
  ``{Consistency checks of grand unified theories},'' {\em Phys. Lett.} {\bf
  B281} (1992)
374--383.

\bibitem{Allanach:2001kg}
B.~C. Allanach, ``Softsusy: A c++ program for calculating supersymmetric
  spectra,'' {\em Comput. Phys. Commun.} {\bf 143} (2002) 305--331,
\href{http://arXiv.org/abs/hep-ph/0104145}{{\tt hep-ph/0104145}}.

\bibitem{Paige:2003mg}
F.~E. Paige, S.~D. Protopopescu, H.~Baer, and X.~Tata, ``Isajet 7.69: A monte
  carlo event generator for p p, anti-p p, and e+ e- reactions,''
\href{http://arXiv.org/abs/hep-ph/0312045}{{\tt hep-ph/0312045}}.

\bibitem{Porod:2003um}
W.~Porod, ``Spheno, a program for calculating supersymmetric spectra, susy
  particle decays and susy particle production at e+ e- colliders,'' {\em
  Comput. Phys. Commun.} {\bf 153} (2003) 275--315,
\href{http://arXiv.org/abs/hep-ph/0301101}{{\tt hep-ph/0301101}}.

\bibitem{ffmssmsc}
A.~Sheplyakov, ``ffmssmsc -- a c++ library for superpartner mass calculation
  and renormalization group analysis of the mssm.'' The source code can be
  obtained form \url{http://theor.jinr.ru/~varg/git/hep/ffmssmsc.git}.

\bibitem{Pierce:1992hg}
D.~Pierce and A.~Papadopoulos, ``Radiative corrections to the higgs boson decay
  rate gamma (h $\to$ z z) in the minimal supersymmetric model,'' {\em Phys.
  Rev.} {\bf D47} (1993) 222--231,
\href{http://arXiv.org/abs/hep-ph/9206257}{{\tt hep-ph/9206257}}.

\bibitem{Bednyakov:2002sf}
A.~Bednyakov, A.~Onishchenko, V.~Velizhanin, and O.~Veretin, ``Two-loop
  $\mathcal{O}(\alpha_s^2)$ mssm corrections to the pole masses of heavy
  quarks,'' {\em Eur. Phys. J.} {\bf C29} (2003) 87--101,
\href{http://arXiv.org/abs/hep-ph/0210258}{{\tt hep-ph/0210258}}.

\bibitem{Bednyakov:2004gr}
A.~Bednyakov and A.~Sheplyakov, ``Two-loop $\mathcal{O}(\alpha_s y^2)$ and
  $\mathcal{O}(y^4)$ mssm corrections to the pole mass of the b-quark,'' {\em
  Phys. Lett.} {\bf B604} (2004) 91--97,
\href{http://arXiv.org/abs/hep-ph/0410128}{{\tt hep-ph/0410128}}.

\bibitem{Tarrach:1980up}
R.~Tarrach, ``The pole mass in perturbative qcd,'' {\em Nucl. Phys.} {\bf B183}
  (1981)
384.

\bibitem{Kronfeld:1998di}
A.~S. Kronfeld, ``The perturbative pole mass in {QCD},'' {\em Phys. Rev.} {\bf
  D58} (1998) 051501,
\href{http://arXiv.org/abs/hep-ph/9805215}{{\tt hep-ph/9805215}}.

\bibitem{Siegel:1979wq}
W.~Siegel, ``Supersymmetric dimensional regularization via dimensional
  reduction,'' {\em Phys. Lett.} {\bf B84} (1979)
193.

\bibitem{Siegel:1980qs}
W.~Siegel, ``Inconsistency of supersymmetric dimensional regularization,'' {\em
  Phys. Lett.} {\bf B94} (1980)
37.

\bibitem{Stockinger:2005gx}
D.~Stockinger, ``Regularization by dimensional reduction: Consistency, quantum
  action principle, and supersymmetry,'' {\em JHEP} {\bf 03} (2005) 076,
\href{http://arXiv.org/abs/hep-ph/0503129}{{\tt hep-ph/0503129}}.

\bibitem{Appelquist:1974tg}
T.~Appelquist and J.~Carazzone, ``Infrared singularities and massive fields,''
  {\em Phys. Rev.} {\bf D11} (1975)
2856.

\bibitem{Beneke:1994sw}
M.~Beneke and V.~M. Braun, ``Heavy quark effective theory beyond perturbation
  theory: Renormalons, the pole mass and the residual mass term,'' {\em Nucl.
  Phys.} {\bf B426} (1994) 301--343,
\href{http://arXiv.org/abs/hep-ph/9402364}{{\tt hep-ph/9402364}}.

\bibitem{Georgi:1994qn}
H.~Georgi, ``Effective field theory,'' {\em Ann. Rev. Nucl. Part. Sci.} {\bf
  43} (1993)
209--252.

\bibitem{Bernreuther:1981sg}
W.~Bernreuther and W.~Wetzel, ``Decoupling of heavy quarks in the minimal
  subtraction scheme,'' {\em Nucl. Phys.} {\bf B197} (1982)
228.

\bibitem{Chetyrkin:2000yt}
K.~G. Chetyrkin, J.~H. Kuhn, and M.~Steinhauser, ``Rundec: A mathematica
  package for running and decoupling of the strong coupling and quark masses,''
  {\em Comput. Phys. Commun.} {\bf 133} (2000) 43--65,
\href{http://arXiv.org/abs/hep-ph/0004189}{{\tt hep-ph/0004189}}.

\bibitem{Bednyakov:2007vm}
A.~V. Bednyakov, ``Running mass of the b-quark in qcd and susy qcd,''
\href{http://arXiv.org/abs/arXiv:0707.0650 [hep-ph]}{{\tt arXiv:0707.0650
  [hep-ph]}}.

\bibitem{Bauer:2008bj}
A.~Bauer, L.~Mihaila, and J.~Salomon, ``{Matching coefficients for $\alpha_s$
  and $m_b$ to $\mathcal{O}(\alpha_s^2)$ in the MSSM},'' {\em JHEP} {\bf 02}
  (2009) 037,
\href{http://arXiv.org/abs/0810.5101}{{\tt 0810.5101}}.

\bibitem{Pierce:1996zz}
D.~M. Pierce, J.~A. Bagger, K.~T. Matchev, and R.-j. Zhang, ``Precision
  corrections in the minimal supersymmetric standard model,'' {\em Nucl. Phys.}
  {\bf B491} (1997) 3--67,
\href{http://arXiv.org/abs/hep-ph/9606211}{{\tt hep-ph/9606211}}.

\bibitem{Amsler:2008zzb}
{\bf Particle Data Group} Collaboration, C.~Amsler {\em et al.}, ``{Review of
  particle physics},'' {\em Phys. Lett.} {\bf B667} (2008)
1.

\bibitem{Baer:2005pv}
H.~Baer, J.~Ferrandis, S.~Kraml, and W.~Porod, ``On the treatment of threshold
  effects in susy spectrum computations,'' {\em Phys. Rev.} {\bf D73} (2006)
  015010,
\href{http://arXiv.org/abs/hep-ph/0511123}{{\tt hep-ph/0511123}}.

\bibitem{Giudice:2004tc}
G.~F. Giudice and A.~Romanino, ``Split supersymmetry,'' {\em Nucl. Phys.} {\bf
  B699} (2004) 65--89,
\href{http://arXiv.org/abs/hep-ph/0406088}{{\tt hep-ph/0406088}}.

\bibitem{Jack:2004ch}
I.~Jack, D.~R.~T. Jones, and A.~F. Kord, ``Snowmass benchmark points and
  three-loop running,'' {\em Ann. Phys.} {\bf 316} (2005) 213--233,
\href{http://arXiv.org/abs/hep-ph/0408128}{{\tt hep-ph/0408128}}.

\bibitem{Harlander:2007wh}
R.~V. Harlander, L.~Mihaila, and M.~Steinhauser, ``Running of $\alpha_s$ and
  $m_b$ in the mssm,''
\href{http://arXiv.org/abs/arXiv:0706.2953 [hep-ph]}{{\tt arXiv:0706.2953
  [hep-ph]}}.

\bibitem{Tkachov:1983pa}
F.~V. Tkachov, ``{EUCLIDEAN ASYMPTOTICS OF FEYNMAN INTEGRALS: BASIC
  NOTIONS},''. IYaI-P-0332.

\bibitem{Tkachov:1984us}
F.~V. Tkachov, ``{ASYMPTOTICS OF EUCLIDEAN FEYNMAN INTEGRALS. 2. ONE LOOP
  CASE},''. IYaI-P-0358.

\bibitem{Smirnov:2002pj}
V.~A. Smirnov, ``Applied asymptotic expansions in momenta and masses,'' {\em
  Springer Tracts Mod. Phys.} {\bf 177} (2002)
1--262.

\bibitem{Haestier:2005ja}
J.~Haestier, S.~Heinemeyer, D.~Stockinger, and G.~Weiglein, ``Electroweak
  precision observables: Two-loop yukawa corrections of supersymmetric
  particles,'' {\em JHEP} {\bf 12} (2005) 027,
\href{http://arXiv.org/abs/hep-ph/0508139}{{\tt hep-ph/0508139}}.

\bibitem{Hempfling:1994ar}
R.~Hempfling and B.~A. Kniehl, ``On the relation between the fermion pole mass
  and ms yukawa coupling in the standard model,'' {\em Phys. Rev.} {\bf D51}
  (1995) 1386--1394,
\href{http://arXiv.org/abs/hep-ph/9408313}{{\tt hep-ph/9408313}}.

\bibitem{Jegerlehner:2003py}
F.~Jegerlehner and M.~Y. Kalmykov, ``The $\mathcal{O}(\alpha \alpha_s)$
  correction to the pole mass of the $t$- quark within the standard model,''
  {\em Nucl. Phys.} {\bf B676} (2004) 365--389,
\href{http://arXiv.org/abs/hep-ph/0308216}{{\tt hep-ph/0308216}}.

\bibitem{Faisst:2004gn}
M.~Faisst, J.~H. Kuhn, and O.~Veretin, ``Pole- versus ms-mass definitions in
  the electroweak theory,'' {\em Phys. Lett.} {\bf B589} (2004) 35--38,
\href{http://arXiv.org/abs/hep-ph/0403026}{{\tt hep-ph/0403026}}.

\bibitem{Castano:1993ri}
D.~J. Castano, E.~J. Piard, and P.~Ramond, ``Renormalization group study of the
  standard model and its extensions. 2. the minimal supersymmetric standard
  model,'' {\em Phys. Rev.} {\bf D49} (1994) 4882--4901,
\href{http://arXiv.org/abs/hep-ph/9308335}{{\tt hep-ph/9308335}}.

\bibitem{Allanach:1999mh}
B.~C. Allanach, A.~Dedes, and H.~K. Dreiner, ``2-loop supersymmetric
  renormalisation group equations including r-parity violation and aspects of
  unification,'' {\em Phys. Rev.} {\bf D60} (1999) 056002,
\href{http://arXiv.org/abs/hep-ph/9902251}{{\tt hep-ph/9902251}}.

\bibitem{Fleischer:1998dw}
J.~Fleischer, F.~Jegerlehner, O.~V. Tarasov, and O.~L. Veretin, ``Two-loop
  {QCD} corrections of the massive fermion propagator,'' {\em Nucl. Phys.} {\bf
  B539} (1999) 671--690,
\href{http://arXiv.org/abs/hep-ph/9803493}{{\tt hep-ph/9803493}}.

\bibitem{Vermaseren:2000nd}
J.~A.~M. Vermaseren, ``New features of form.,''
\href{http://arXiv.org/abs/math-ph/0010025}{{\tt math-ph/0010025}}.

\bibitem{FeynArts}
T.~Hahn, {\em FeynArts 3.2. User's guide}.

\bibitem{Hahn:2001rv}
T.~Hahn and C.~Schappacher, ``The implementation of the minimal supersymmetric
  standard model in feynarts and formcalc,'' {\em Comput. Phys. Commun.} {\bf
  143} (2002) 54--68,
\href{http://arXiv.org/abs/hep-ph/0105349}{{\tt hep-ph/0105349}}.

\bibitem{DBLP:journals/corr/cs-SC-0004015}
C.~Bauer, A.~Frink, and R.~Kreckel, ``Introduction to the ginac framework for
  symbolic computation within the c++ programming language,'' {\em CoRR} {\bf
  cs.SC/0004015} (2000).

\bibitem{Carena:1999py}
M.~S. Carena, D.~Garcia, U.~Nierste, and C.~E.~M. Wagner, ``Effective
  lagrangian for the anti-t b h+ interaction in the mssm and charged higgs
  phenomenology,'' {\em Nucl. Phys.} {\bf B577} (2000) 88--120,
\href{http://arXiv.org/abs/hep-ph/9912516}{{\tt hep-ph/9912516}}.

\bibitem{Allanach:2002nj}
B.~C. Allanach {\em et al.}, ``The snowmass points and slopes: Benchmarks for
  susy searches,''
\href{http://arXiv.org/abs/hep-ph/0202233}{{\tt hep-ph/0202233}}.

\bibitem{Evans:2008zzb}
L.~Evans, (ed.~) and P.~Bryant, (ed.~), ``{LHC Machine},'' {\em JINST} {\bf 3}
  (2008)
S08001.

\bibitem{bubblesII}
A.~Sheplyakov, ``bubblesii, a c++ library for analytical and numerical
  evaluation of 2-loop vacuum integrals.'' The source code can be obtained form
  \url{http://theor.jinr.ru/~varg/dist}.

\bibitem{Kazakov:2004mr}
D.~I. Kazakov, ``Beyond the standard model,'' CERN: 2006-003,
\href{http://arXiv.org/abs/hep-ph/0411064}{{\tt hep-ph/0411064}}.

\end{thebibliography}
\end{document}